\documentclass[aps,twocolumn,prb,showpacs,floatfix,superscriptaddress,citeautoscript]{revtex4-1}

\usepackage{epsfig,amsmath,amssymb}
\usepackage{graphics} 
\usepackage{subfigure}
\usepackage{color}
\usepackage{multirow}
\usepackage{mathrsfs}
\usepackage{natbib}

\begin{document}

\title
{
Phase diagram of a frustrated spin-$\frac{1}{2}$ $J_{1}$--$J_{2}$ $XXZ$ model on the honeycomb lattice
}

\author
{P.~H.~Y.~Li}
\affiliation
{School of Physics and Astronomy, Schuster Building, The University of Manchester, Manchester, M13 9PL, UK}

\author
{R.~F.~Bishop}
\affiliation
{School of Physics and Astronomy, Schuster Building, The University of Manchester, Manchester, M13 9PL, UK}

\author
{C.~E.~Campbell}
\affiliation
{School of Physics and Astronomy, University of Minnesota, 116 Church Street SE, Minneapolis, Minnesota 55455, USA}

\begin{abstract}
  We study the zero-temperature ($T=0$) ground-state (GS) properties of
  a frustrated spin-half $J^{XXZ}_{1}$--$J^{XXZ}_{2}$ model on the
honeycomb lattice with nearest-neighbor and next-nearest-neighbor
interactions with exchange couplings $J_{1}>0$ and $J_{2} \equiv
\kappa J_{1}>0$, respectively, using the coupled cluster method.  Both
interactions are of the anisotropic $XXZ$ type.  We present the
$T=0$ GS phase diagram of the model in the ranges $0 \leq \Delta \leq
1$ of the spin-space anisotropy parameter and $0 \leq \kappa \leq 1$
of the frustration parameter.  A possible quantum spin-liquid region is identified.
\end{abstract}


\maketitle

Frustrated spin-half ($s=\frac{1}{2}$) antiferromagnets with
nearest-neighbor (NN) $J_{1}>0$ and competing next-nearest-neighbor
(NNN) $J_{2}>0$ exchange couplings on the honeycomb lattice have
attracted a great deal of interest in recent years.  These have
included the two specific cases where both couplings have either an
isotropic Heisenberg ($XXX$) form (see, e.g., Refs.\
\onlinecite{Rastelli:1979_honey,Fouet:2001_honey,Mulder:2010_honey,
  Ganesh:2011_honey,Clark:2011_honey,Albuquerque:2011_honey,
  Mosadeq:2011_honey,Oitmaa:2011_honey,Mezzacapo:2012_honey,Li:2012_honey_full,Bishop:2012_honeyJ1-J2,RFB:2013_hcomb_SDVBC,Ganesh:2013_honey_J1J2mod-XXX,Zhu:2013_honey_J1J2mod-XXZ,Gong:2013_J1J2mod-XXX,Yu:2014_honey_J1J2mod-XXZ},
and references cited therein) or an isotropic $XY$ ($XX$) form (see,
e.g., Refs.\
\onlinecite{Varney:2011_honey_XY,Varney:2012_honey_XY,Zhu:2013_honey_XY,Carrasquilla:2013_honey_XY,Ciolo:2014_honey_XY,Oitmaa:2014_honey_XY,Bishop:2014_honey_XY}).
Although the classical ($s \rightarrow \infty$) versions of these two
models have identical zero-temperature ($T=0$) ground-state (GS) phase
diagrams \cite{Rastelli:1979_honey,Fouet:2001_honey}, their
$s=\frac{1}{2}$ counterparts differ in significant ways.  Furthermore,
there is not yet a complete consensus on the GS phase orderings for
either model in the range $0 \leq \kappa \leq 1$ of the frustration
parameter $\kappa \equiv J_{2}/J_{1}$.

Whereas both classical ($s \rightarrow \infty$) models have
N\'{e}el ordering for $\kappa < \kappa_{{\rm cl}} = \frac{1}{6}$, the spin-$\frac{1}{2}$ models both seem to retain N\'{e}el order out
to larger values $\kappa_{c_{1}} \approx 0.2$, consistent with the
fact that quantum fluctuations generally favor collinear
over noncollinear ordering.  The degenerate family of spiral states
that form the classical GS phase for all values
$\kappa > \kappa_{{\rm cl}}$ is very fragile against quantum fluctuations, and there is broad agreement that neither $s=\frac{1}{2}$ model has a stable GS phase
with spiral ordering for any value of $\kappa$ in the range $0 \leq
\kappa \leq 1$.

The most interesting, and also most uncertain, regime for both
$s=\frac{1}{2}$ models is when $0.2 \lesssim \kappa \lesssim 0.4$.  For
the $XXX$ model the N\'{e}el order that exists for $\kappa <
\kappa_{c_{1}} \approx 0.2$ is predicted by different methods to give
way either to a GS phase with plaquette valence-bond crystalline
(PVBC) order \cite{Albuquerque:2011_honey,
  Mosadeq:2011_honey,Li:2012_honey_full,Bishop:2012_honeyJ1-J2,RFB:2013_hcomb_SDVBC,Ganesh:2013_honey_J1J2mod-XXX,Zhu:2013_honey_J1J2mod-XXZ} or to a
quantum spin-liquid (QSL) state \cite{Clark:2011_honey,Mezzacapo:2012_honey,Gong:2013_J1J2mod-XXX,Yu:2014_honey_J1J2mod-XXZ} in the range $\kappa_{c_{1}} < \kappa < \kappa_{c_{2}}
\approx 0.4$.  By contrast, for the $XX$ model the
N\'{e}el $xy$ planar [N(p)] ordering that exists for $\kappa <
\kappa_{c_{1}}$ is predicted by different methods to yield either to a GS
phase with N\'{e}el $z$-aligned [N($z$)] order
\cite{Zhu:2013_honey_XY,Bishop:2014_honey_XY} or to a QSL state \cite{Varney:2011_honey_XY,Carrasquilla:2013_honey_XY} in a
corresponding range $\kappa_{c_{1}} < \kappa < \kappa_{c_{2}}$.  There is broad
agreement for both models that for $(1 >)$ $\kappa > \kappa_{c_{2}}$
there is a strong competition to form the GS phase between states with
collinear N\'{e}el-II $xy$ planar [N-II(p)] and staggered dimer
valence-bond crystalline (SDVBC) forms of order, which lie very close
in energy.  The (threefold-degenerate) N\'{e}el-II states, which break
the lattice rotational symmetry, are ones in which NN pairs of spins
are parallel along one of the three equivalent honeycomb directions
and antiparallel along the other two.  Some methods favor a further
quantum critical point (QCP) at $\kappa_{c_{3}} > \kappa_{c_{2}}$, at
which a transition occurs between GS phases with N-II(p) ordering for
$\kappa_{c_{2}} < \kappa < \kappa_{c_{3}}$ and SDVBC ordering for
$\kappa > \kappa_{c_{3}}$.

\begin{figure*}[!tbh]
\begin{center}
\mbox{
\subfigure[]{\scalebox{0.28}{\includegraphics{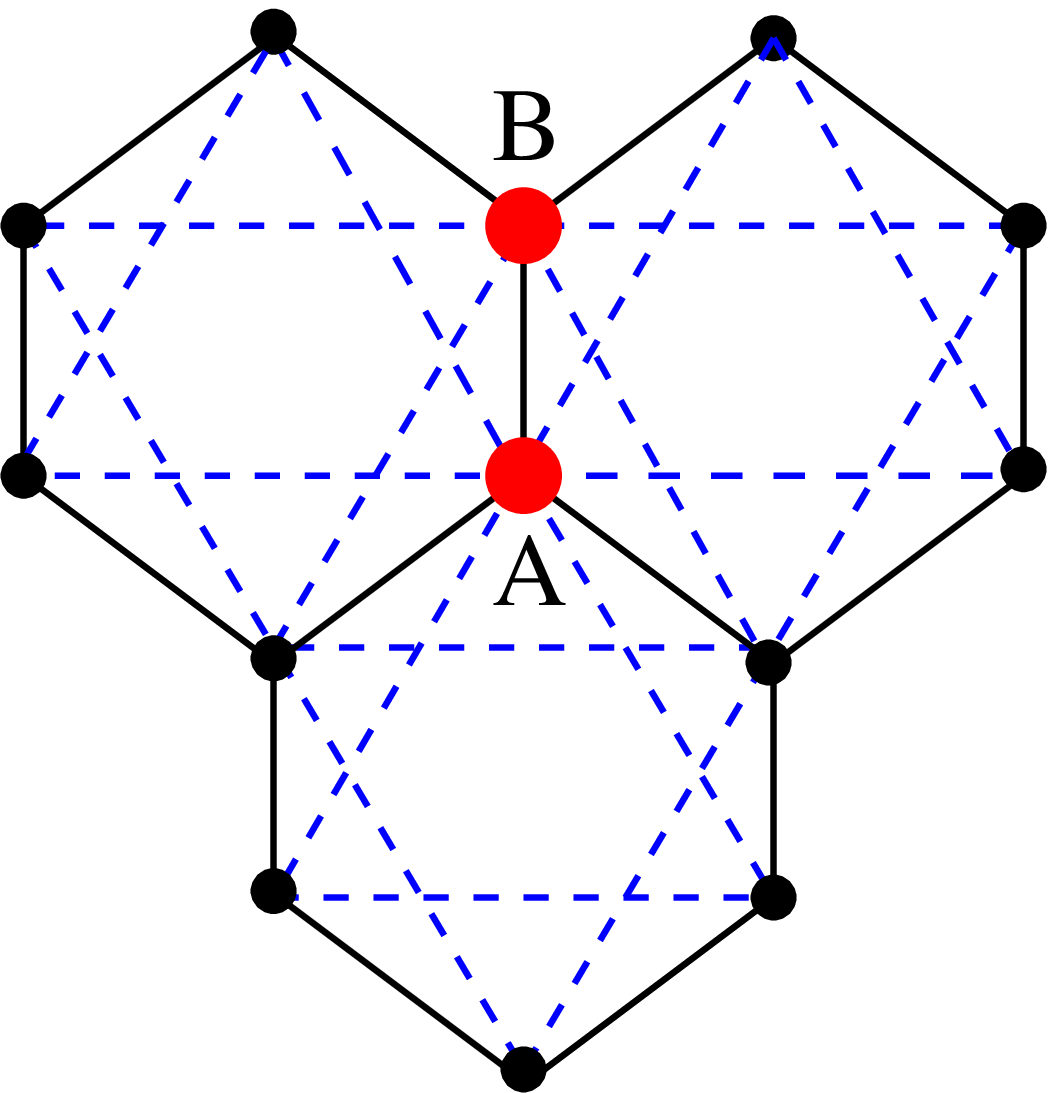}}}
\quad
\subfigure[]{\scalebox{0.2}{\includegraphics{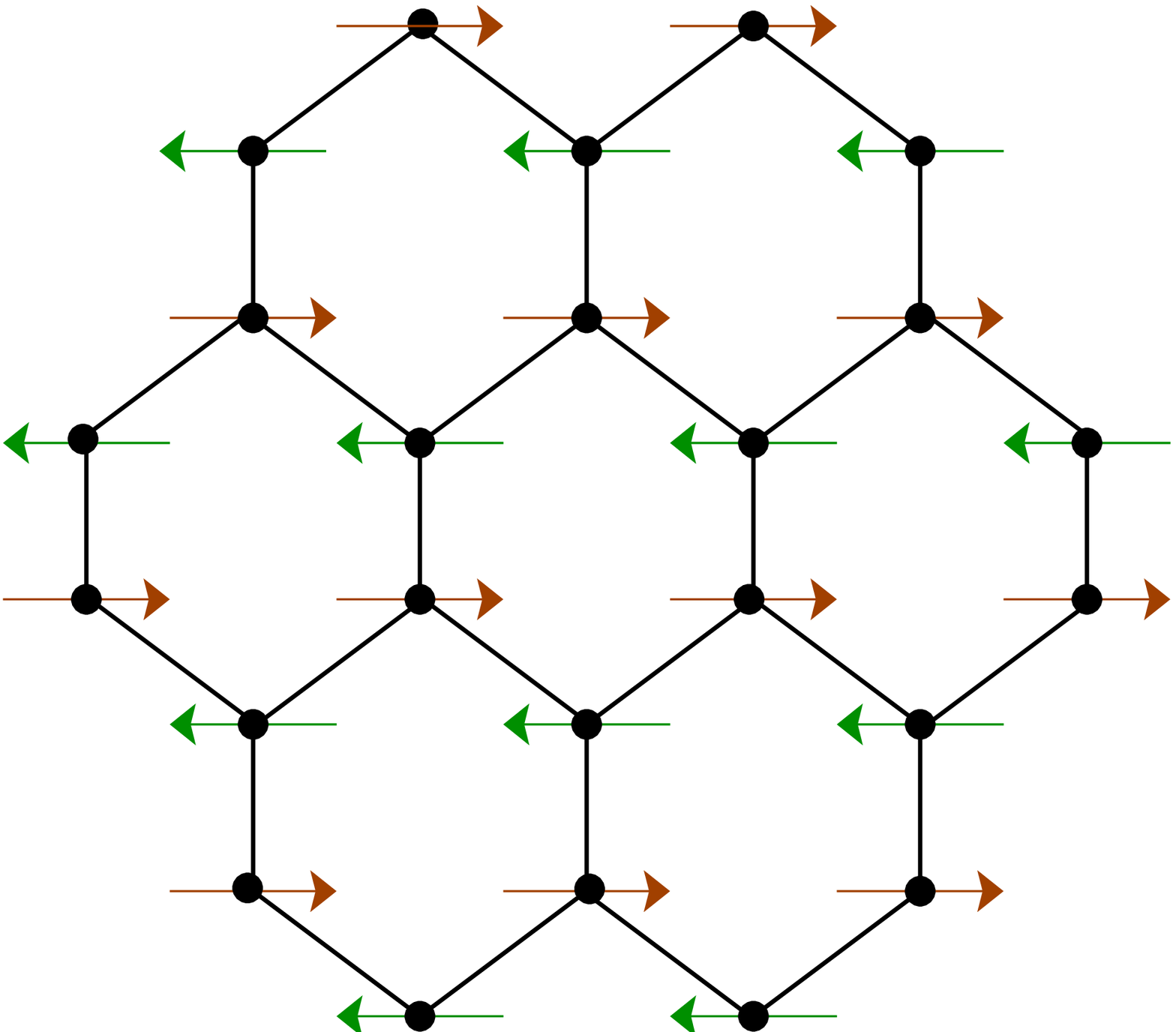}}}
\quad
\subfigure[]{\scalebox{0.2}{\includegraphics{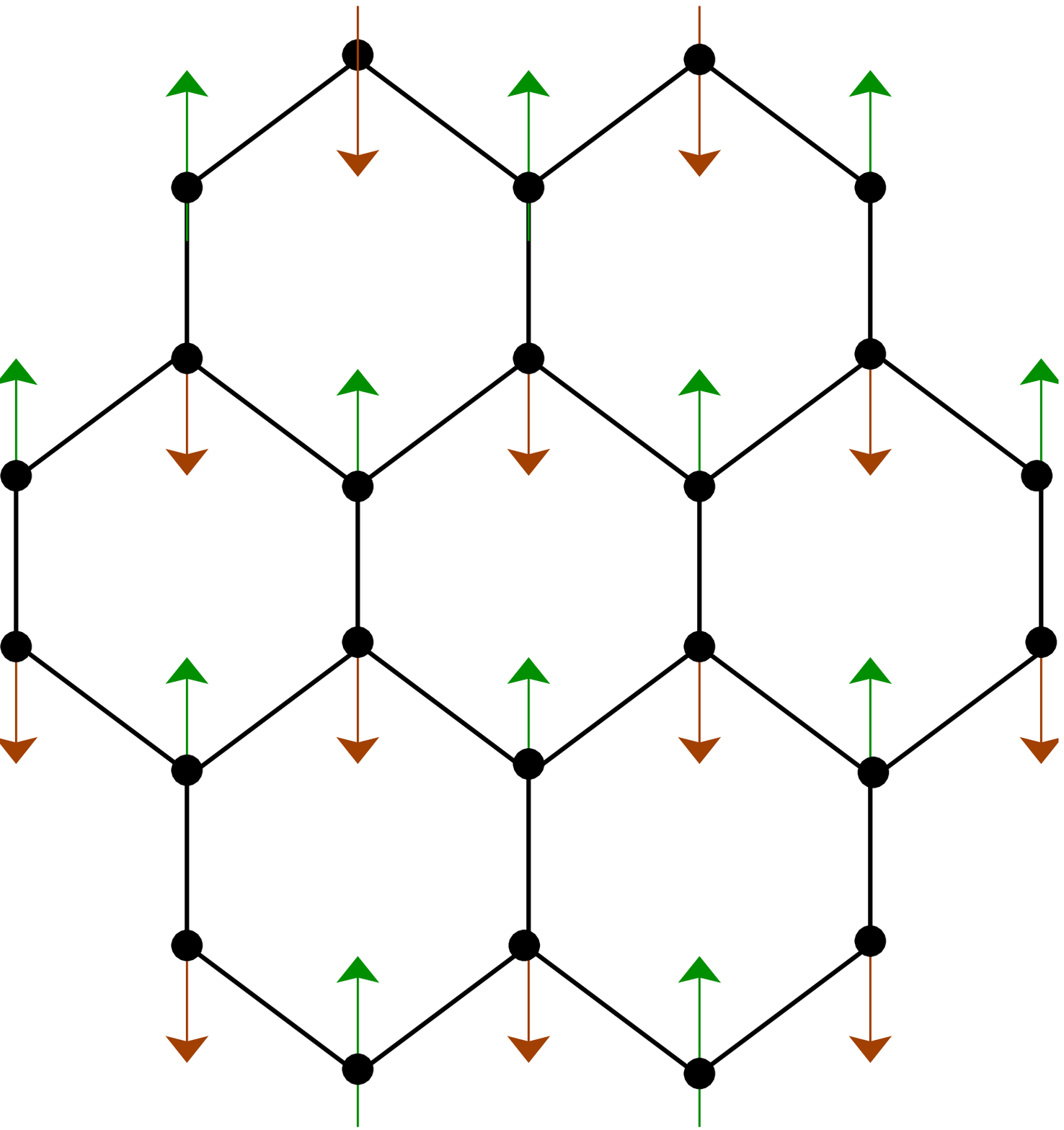}}}
\quad
\subfigure[]{\scalebox{0.2}{\includegraphics{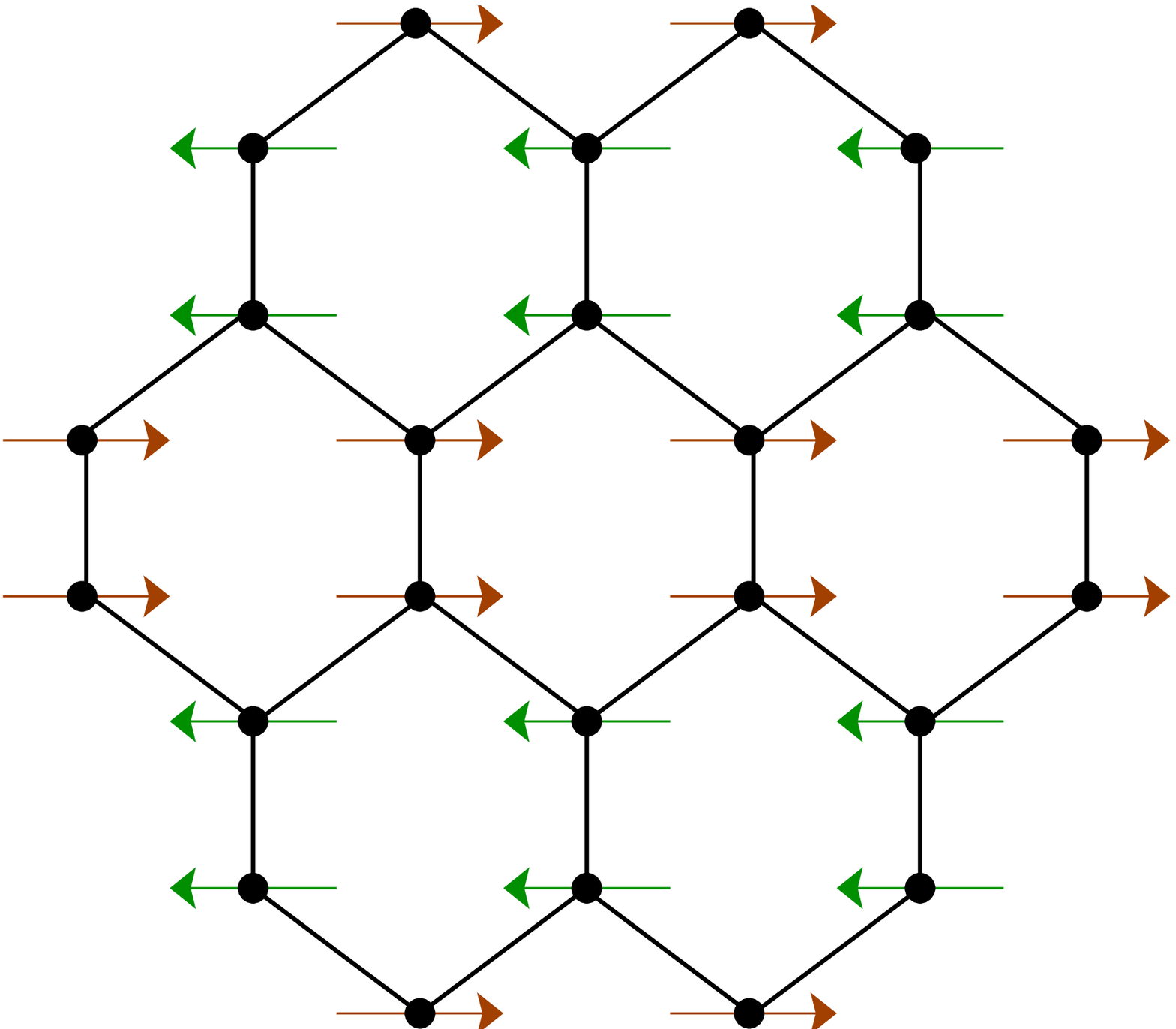}}}
}
\caption{(color online)  The $J^{XXZ}_{1}$--$J^{XXZ}_{2}$ model on the honeycomb lattice,
  showing (a) the bonds ($J_{1} \equiv$ ----- ; $J_{2} \equiv \textcolor{blue}{- - -}$) and the two sites (\textcolor{red}{{\Large $\bullet$}}) A and B of the unit cell; (b) the N\'{e}el planar, N(p), state; (c) the N\'{e}el $z$-aligned, N($z$), state; and (d) the N\'{e}el-II planar, N-II(p), state.  The arrows represent the directions of the spins located on lattice sites \textbullet.}
\label{model}
\end{center}
\end{figure*}  

The intriguing differences between the two models motivate us to consider the so-called
$J^{XXZ}_{1}$--$J^{XXZ}_{2}$ model that interpolates between them.
It is shown schematically in Fig.\ \ref{model}(a) and is described by the Hamiltonian,
\begin{eqnarray}
H & = & J_{1}\sum_{\langle i,j \rangle}(s^{x}_{i}s^{x}_{j}+s^{y}_{i}s^{y}_{j}+\Delta s^{z}_{i}s^{z}_{j}) + \nonumber \\
  &  & J_{2}\sum_{\langle\langle i,k \rangle\rangle}(s^{x}_{i}s^{x}_{k}+s^{y}_{i}s^{y}_{k}+\Delta s^{z}_{i}s^{z}_{k})\,, \label{H_honey_XXZ}
\end{eqnarray}
where $\langle i,j \rangle$ and $\langle\langle i,k \rangle\rangle$
indicate NN and NNN pairs of spins, respectively, and ${\bf
  s}_{i}=(s^{x}_{i},s^{y}_{i},s^{z}_{i}$) is the spin operator on
lattice site $i$.  We shall study the $T=0$ GS phase diagram for the
spin-$\frac{1}{2}$ Hamiltonian of Eq.\ (\ref{H_honey_XXZ}) on the
honeycomb lattice in the range $0 \leq \Delta \leq 1$ of the spin
anisotropy parameter that spans from the $XX$ model (with $\Delta =
0$) to the $XXX$ model (with $\Delta = 1$), and in the range $0 \leq
\kappa \leq 1$ of the frustration parameter.  Henceforth we put $J_{1}
\equiv 1$ to set the overall energy scale.  We note that both exact
diagonalization (ED) of small finite lattices and density-matrix
renormalization group (DMRG) studies of the $XX$ model in particular
find it especially difficult to distinguish the N-II(p) and SDVBC
phases in the regime $\kappa > \kappa_{c_{2}}$ in the thermodynamic
limit, $N \rightarrow \infty$, where $N$ is the number of lattice
sites.  For this reason it is particularly suitable to use a
size-extensive method such as the coupled cluster method (CCM) that works from the outset in the $N \rightarrow \infty$
limit.

We first describe some key features of the CCM and refer
the reader to Refs.\
\onlinecite{Bishop:2012_honeyJ1-J2,RFB:2013_hcomb_SDVBC,Bishop:2014_honey_XY,Bishop:1987_ccm,Arponen:1991_ccm,Bishop:1991_TheorChimActa_QMBT,Bishop:1998_QMBT_coll,Bishop:1991_XXZ_PRB44,Zeng:1998_SqLatt_TrianLatt,Fa:2004_QM-coll,Bi:2008_PRB_J1xxzJ2xxz}
for more details.  Any CCM calculation starts with the choice of a
suitable model (or reference) state $|\Phi\rangle$.  Here we use each
of the N(p), N($z$), and N-II(p) states shown schematically in Figs.\
\ref{model}(b)--(d).  
In order to treat each lattice site on an equal footing we 
passively rotate each spin in each model state, so that in its own
local spin-coordinate frame it points downwards (i.e., along the local
negative $z$ axis).  In these local spin coordinates every model state
thus takes the universal form
$|\Phi\rangle=|\downarrow\downarrow\downarrow\cdots\downarrow\rangle$
and the Hamiltonian has to be rewritten accordingly.  The exact GS
energy eigen-ket, $|\Psi\rangle$, with $H|\Psi\rangle=E|\Psi\rangle$, is now expressed in the exponentiated form, $|\Psi\rangle=e^{S}|\Phi\rangle$,
where the creation correlation operator $S$ is written as
$S=\sum_{I\neq 0}{\cal S}_{I}C^{+}_{I}$, with $C^{+}_{0}\equiv 1$, the
identity operator.  The corresponding GS energy eigen-bra,
$\langle\tilde{\Psi}|$, where $\langle\tilde{\Psi}|H = E\langle\tilde{\Psi}|$,
is written as $\langle\tilde{\Psi}|=\langle\Phi|\tilde{S}e^{-S}$, where
$\tilde{S}=1+\sum_{I\neq 0}\tilde{{\cal S}}_{I}C^{-}_{I}$, and
$C^{-}_{I} \equiv (C^{+}_{I})^{\dagger}$.  The states obey the
normalization conditions $\langle\tilde{\Psi}|\Psi\rangle =
\langle{\Phi}|\Psi\rangle = \langle{\Phi}|\Phi\rangle = 1$, and the relations, $\langle\Phi|C^{+}_{I} = 0 =
C^{-}_{I}|\Phi\rangle,\,\forall I \neq 0$, which ensure that
$|\Phi\rangle$ is a fiducial vector with respect to the complete set
of multispin creation operators $\{C^{+}_{I}\}$.  In the local
spin-coordinate frames, $C^{+}_{I}$ also takes a universal form,
$C^{+}_{I} \rightarrow s^{+}_{l_{1}}s^{+}_{l_{2}}\cdots
s^{+}_{l_{n}}$, a product of single-spin raising operators, $s^{+}_{l}
\equiv s^{x}_{l}+is^{y}_{l}$, where the set-index $I \rightarrow
\{l_{1},l_{2},\cdots , l_{n};\; n=1,2,\cdots , 2sN\}$.  The set of
multispin correlation coefficients $\{{\cal S}_{I},{\tilde{\cal
    S}}_{I}\}$ is determined by requiring that the energy expectation
value $\bar{H}=\bar{H}\{{\cal S}_{I},{\tilde{\cal S}_{I}}\}\equiv
\langle\Phi|\tilde{S}e^{-S}He^{S}|\Phi\rangle$ is a minimum.  The GS
magnetic order parameter is defined as $M \equiv
-\frac{1}{N}\langle\tilde{\Psi}|\sum_{k=1}^{N}s^{z}_{k}|\Psi\rangle$,
the average local on-site magnetization, with respect to the local
(rotated) spin coordinates.

The {\it only} approximation now made in the CCM is to truncate the
set of indices $\{I\}$ in the expansions of the correlation operators
$S$ and $\tilde{S}$.  We use here the well-studied
(lattice-animal-based subsystem) LSUB$m$ scheme
\cite{Bishop:2012_honeyJ1-J2,RFB:2013_hcomb_SDVBC,Bishop:2014_honey_XY,Bishop:1991_XXZ_PRB44,Zeng:1998_SqLatt_TrianLatt,Bi:2008_PRB_J1xxzJ2xxz,Fa:2004_QM-coll}
in which, at the $m$th level of approximation, one retains all
multispin-flip configurations $\{I\}$ defined over no more than $m$
contiguous lattice sites.  Such cluster configurations are
defined to be contiguous if every site is NN to
at least one other.  The number, $N_{f}$, of such fundamental
configurations is reduced by exploiting the space- and point-group
symmetries and any conservation laws that pertain to the Hamiltonian
and the model state being used.  Even so, $N_{f}$ increases rapidly
with increasing LSUB$m$ truncation index $m$, and it becomes necessary
to use massive parallelization together with supercomputing resources
\cite{Zeng:1998_SqLatt_TrianLatt,ccm}.  For example, we have finally
$N_{f}=818300$ for the N-II(p) reference state at the LSUB12 level.

Finally, we extrapolate the ``raw'' LSUB$m$ results to the
limit $m \rightarrow \infty$ where the CCM becomes exact.  For the GS
energy per spin, $e \equiv E/N$, we use the well-tested
extrapolation scheme
\cite{Bishop:2012_honeyJ1-J2,RFB:2013_hcomb_SDVBC,Bishop:2014_honey_XY,Bi:2008_PRB_J1xxzJ2xxz,Fa:2004_QM-coll}
$e(m) = e_{0}+e_{1}m^{-2}+e_{2}m^{-4}$, where results with
$m=\{6,8,10,12\}$ are used for the N(p) and N-II(p) states used as
model state, and with $m=\{4,6,8,10\}$ for the N($z$) state.  For the
magnetic order parameter of systems near a QCP an appropriate
extrapolation rule is the ``leading power-law'' scheme
\cite{RFB:2013_hcomb_SDVBC,Bishop:2014_honey_XY},
$M(m)=c_{0}+c_{1}(1/m)^{c_{2}}$, which we use here for the LSUB$m$
results based on the N($z$) state with $m=\{4,6,8,10\}$.  An
alternative well-tested scheme for systems with strong frustration or
where the order in question is zero or close to zero
\cite{Bishop:2012_honeyJ1-J2,RFB:2013_hcomb_SDVBC,Bishop:2014_honey_XY}
is $M(m)=d_{0}+d_{1}m^{-1/2}+d_{2}m^{-3/2}$, when the leading exponent
$c_{2}$ above has been empirically found to be close to 0.5, as is the
case here for results based on both the N(p) and N-II(p) model states with $m=\{6,8,10,12\}$.

\begin{figure}
  \includegraphics[angle=270,width=9cm]{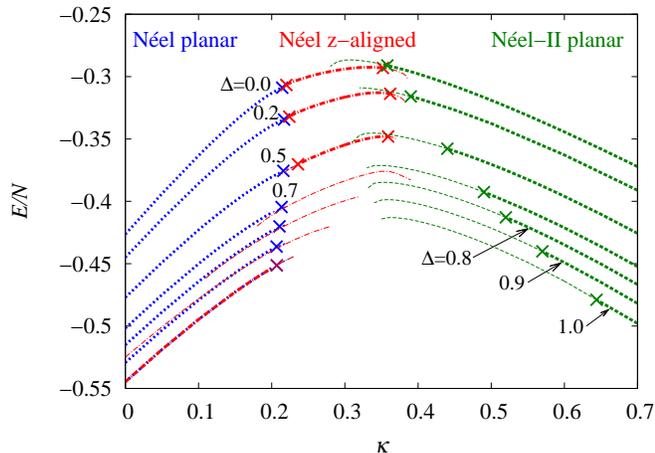}
  \caption{(color online) The GS energy per spin $E/N$ versus the
    frustration parameter $\kappa \equiv J_{2}/J_{1}$ for the
    spin-$\frac{1}{2}$ $J^{XXZ}_{1}$--$J^{XXZ}_{2}$ model on the
    honeycomb lattice (with $J_{1}=1$), for various values of the
    anisotropy parameter $\Delta = 0.0, 0.2, 0.5, 0.7, 0.8, 0.9, 1.0$
    (from top to bottom, respectively).  We show extrapolated CCM
    LSUB$\infty$ results (see text for details) based on the N\'{e}el
    planar, N\'{e}el $z$-aligned, and N\'{e}el-II planar model states, respectively.  The times ($\times$) symbols mark the
    points where the respective extrapolations for the order parameter
    have $M \rightarrow 0$, and the unphysical portions of the
    solutions are shown by thinner lines (see text for details).}
\label{E}
\end{figure}
%%%%%%%%%%%%%
\begin{figure}[t]
  \includegraphics[angle=270,width=9cm]{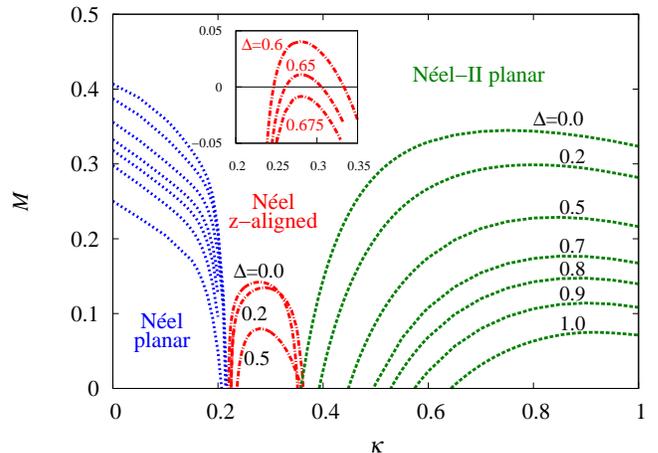}
  \caption{(color online) The GS magnetic order parameter $M$ versus the frustration parameter $\kappa \equiv J_{2}/J_{1}$ for the spin-$\frac{1}{2}$ $J^{XXZ}_{1}$--$J^{XXZ}_{2}$ model on the honeycomb lattice (with $J_{1}>0$) for various values of the anisotropy parameter $\Delta=0.0,0.2,0.5,0.7,0.8,0.9,1.0$ (from top to bottom, respectively).  We show extrapolated CCM LSUB$\infty$ results (see text for details) based on the N\'{e}el planar, N\'{e}el $z$-aligned, and N\'{e}el-II planar states as CCM model states, respectively.}
\label{M}
\end{figure}

Our corresponding extrapolated (LSUB$\infty$) results for the GS
energy per spin and the order parameter are shown in Figs.\ \ref{E}
and \ref{M}.  In each case we present three curves for
each value of the anisotropy parameter $\Delta$ shown, based in turn
on the N(p), N($z$), and N-II(p) model states.  All of the curves in
Fig.\ \ref{E} display termination points, {\it viz.}, an upper one for
the N(p) curves, a lower one for the N-II(p) curves, and one of each type
for the N($z$) curves.  In each case they correspond to the points of
the respective LSUB$m$ approximations with the highest value of $m$,
beyond which real solutions for $\{{\cal S}_{I}\}$ cease to exist.
Such termination points of LSUB$m$ solutions are manifestations
of a corresponding QCP in the system, beyond which the order
associated with the model state under study melts (see, e.g., Refs.\
\onlinecite{Bishop:2012_honeyJ1-J2,RFB:2013_hcomb_SDVBC,Bishop:2014_honey_XY,Fa:2004_QM-coll}).
We find that as the index $m$ is increased the range of values
of $\kappa$ for which the respective LSUB$m$ equations have real
solutions becomes narrower, such that as $m \rightarrow \infty$ each
termination point becomes the corresponding exact QCP.  Real LSUB$m$
solutions with a finite value of $m$ can thus also exist in regions
where the corresponding order is destroyed (viz., where $M<0$).  We
show in Fig.\ \ref{E} by times ($\times$) symbols those points on the
respective curves where $M=0$ (as determined from the corresponding
extrapolated LSUB$\infty$ values shown in Fig.\ \ref{M}).  We also
denote in Fig.\ \ref{E} by thinner lines those portions of the curves
which are ``unphysical'' in the sense that $M<0$, as opposed to the
corresponding ``physical'' regions where $M>0$, which are denoted by
the thicker portions.

Figures \ref{E} and \ref{M} show that:
(a) N(p) order is present below a lower critical value $0 < \kappa <
\kappa_{c_{l}}(\Delta)$, for all values of $\Delta$, where
$\kappa_{c_{l}}(\Delta) \approx 0.21$; (b) N($z$) order is present
within a relatively narrow range of values around $\kappa \approx 0.3$
for all values $\Delta \lesssim 0.66$ and is absent for $\Delta \gtrsim
0.66$; (c) N-II(p) order is present above some upper critical value,
$(1 >)$ $\kappa > \kappa_{c_{u}}(\Delta)$, where
$\kappa_{c_{u}}(\Delta)$ increases monotonically with $\Delta$, (d)
whereas the GS phases with N(p) and N($z$) order present seem to meet
at $\kappa_{c_{l}}(0)$ for $\Delta=0$, a very narrow region of a GS
phase with neither of these orderings opens up between them as
$\Delta$ is increased; and (e) similarly, whereas the GS phases with
N($z$) and N-II(p) order seem to meet at $\kappa_{c_{u}}(0)$ for
$\Delta=0$, a phase with neither order opens between them as $\Delta$
is increased.

\begin{figure}[h]
\begin{center}
\mbox{
%\raisebox{-4.5cm}{
%\raisebox{-4cm}{
%\raisebox{-3.7cm}{
\subfigure{\includegraphics[width=6cm,height=6cm,angle=270]{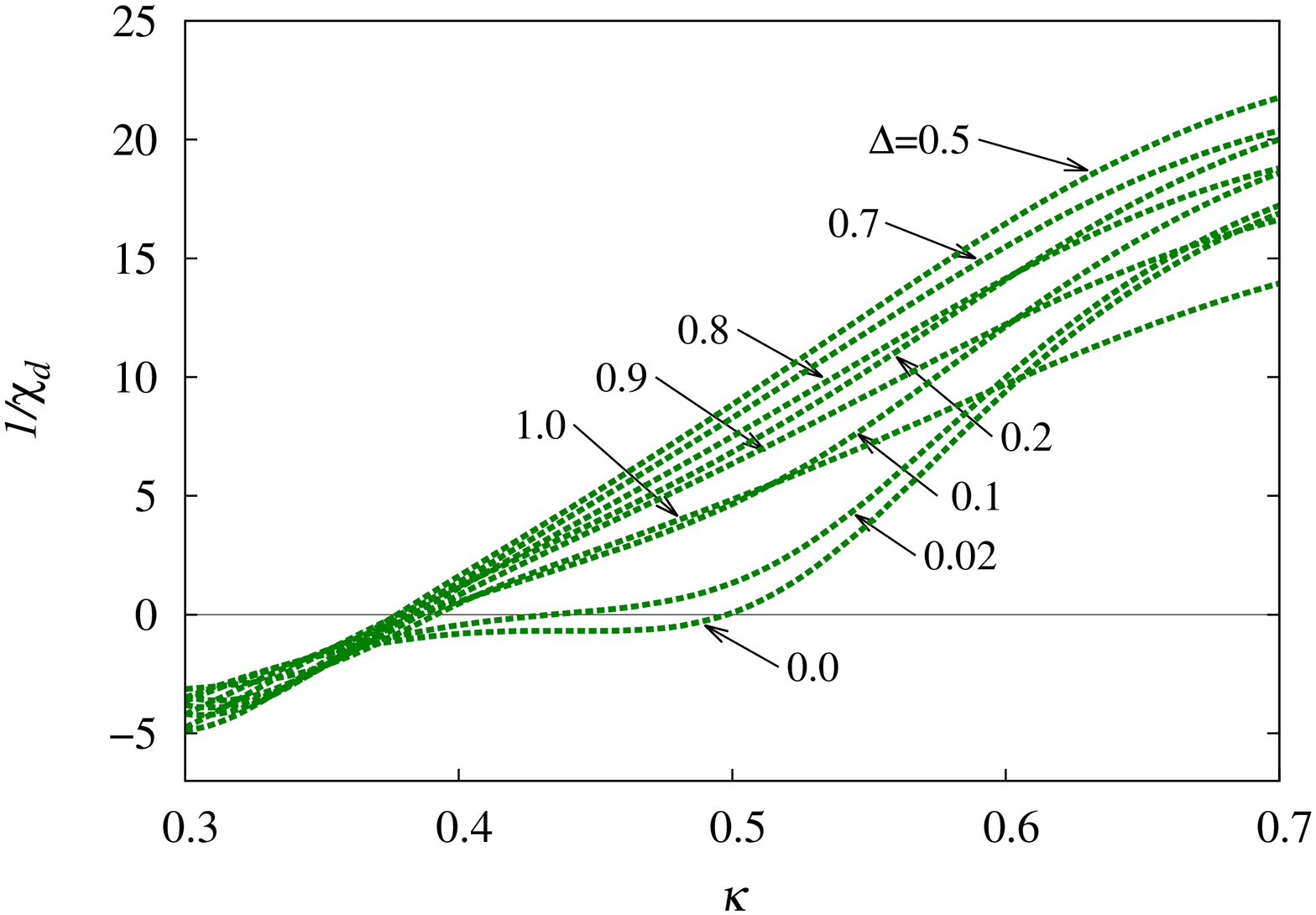}}
\raisebox{-3.5cm}{
\subfigure{\includegraphics[width=2.2cm,height=2.2cm]{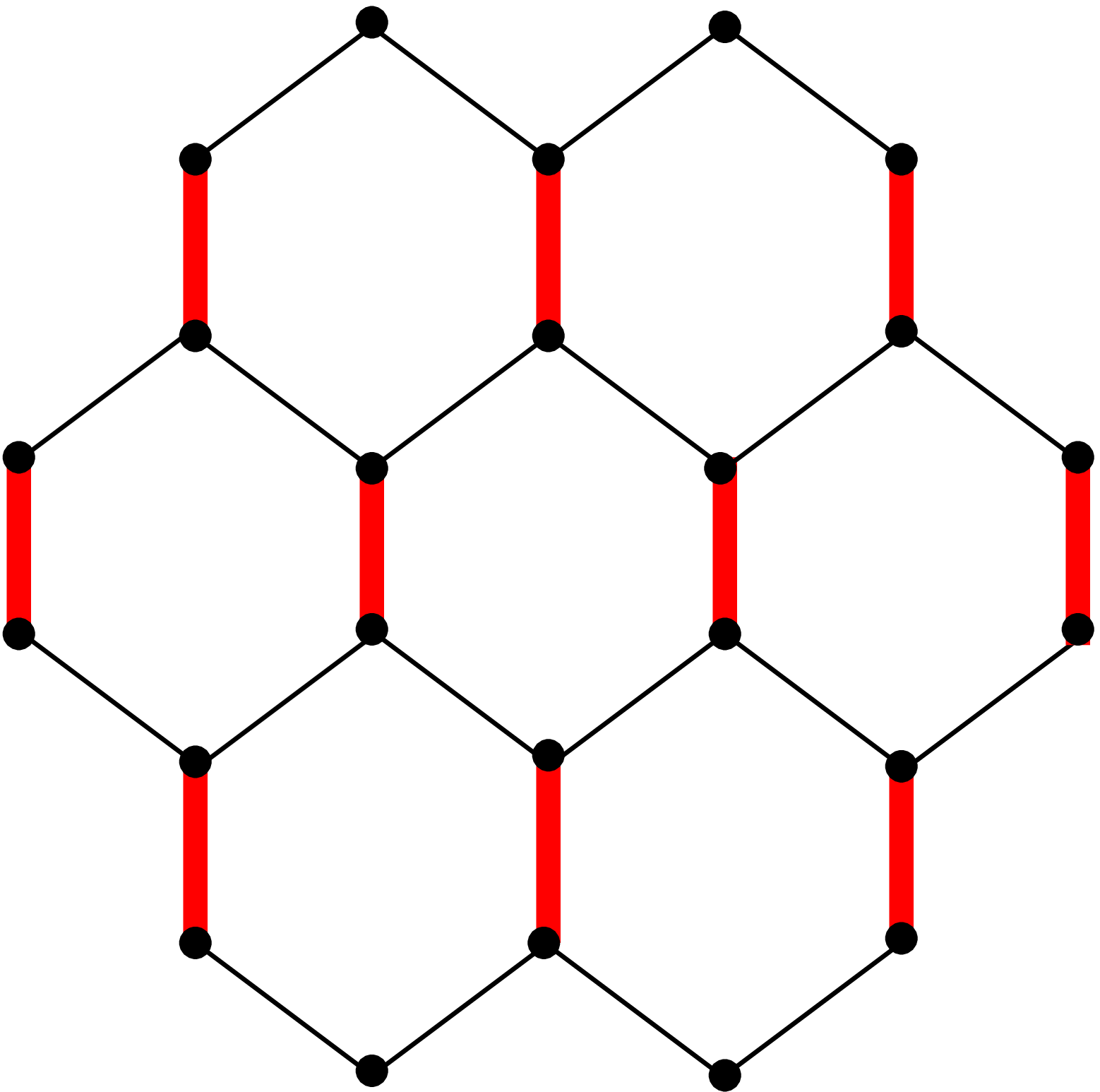}}
}
}
\caption{(color online) Left: The inverse staggered dimer susceptibility, $1/\chi_{d}$, versus the frustration parameter, $\kappa \equiv J_{2}/J_{1}$, for the
  spin-$\frac{1}{2}$ $J^{XXZ}_{1}$--$J^{XXZ}_{2}$ model on the honeycomb lattice (with $J_{1}=1$) for various values of the anisotropy parameter $\Delta$.  We show extrapolated CCM LSUB$\infty$ results (see text for details) based on the N\'{e}el-II planar state as CCM model state.  Right: The field $F \rightarrow
  \delta\; \hat{O}_{d}$ for the staggered dimer susceptibility,
  $\chi_{d}$.  Thick (red) and thin (black) lines correspond
  respectively to strengthened and unaltered NN exchange couplings,
  where $\hat{O}_{d} = \sum_{\langle i,j \rangle} a_{ij}(s^{x}_{i}{s}^{x}_{j}+s^{y}_{i}{s}^{y}_{j}+ \Delta s^{z}_{i}{s}^{z}_{j})$, and the sum runs over all NN
  bonds, with $a_{ij}=+1$ and 0 for thick (red) lines and thin
  (black) lines respectively.}
\label{X_SDVBC}
\end{center}
\end{figure}

From our previous results at $\Delta=0$
\cite{Bishop:2014_honey_XY} and $\Delta=1$
\cite{Bishop:2012_honeyJ1-J2,RFB:2013_hcomb_SDVBC} and those of
others, a possible phase for that mentioned under item (e)
above is one with SDVBC ordering.  A convenient way to test
for the susceptibility of a candidate GS phase built on a specific CCM
model state is to consider its response to an imposed field operator,
$F=\delta\hat{O}$, added to our Hamiltonian of Eq.\ (\ref{H_honey_XXZ}),
where $\delta$ is a (positive) infinitesimal and the operator
$\hat{O}$ now promotes SDVBC order
($\hat{O}_{d}$), as illustrated in Fig.\ \ref{X_SDVBC}.
\begin{figure}[h]
  \includegraphics[angle=270,width=9cm]{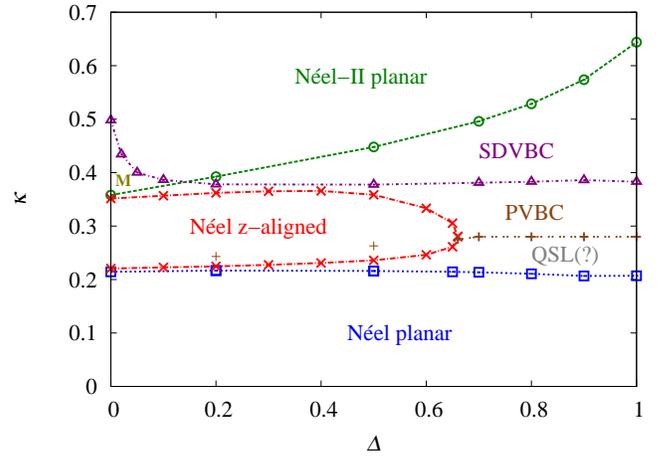}
  \caption{(color online) Phase diagram for the spin-$\frac{1}{2}$
    $J^{XXZ}_{1}$--$J^{XXZ}_{2}$ model on the honeycomb lattice (with
    $J_{1}>0$ and $\kappa \equiv J_{2}/J_{1}>0$) in the window $0 \leq \kappa \leq 1$ and $0 \leq \Delta \leq 1$, as obtained by a CCM analysis.  The phase in the region marked ``M'' has both SDVBC and N\'{e}el-II planar order.  See text for details.}
\label{phase}
\end{figure}
The corresponding perturbed energy per spin, $e(\delta) \equiv
E(\delta)/N$, is calculated at various CCM LSUB$m$ orders of
approximation and used to calculate the respective susceptibility,
$\chi \equiv -\partial^2e(\delta)/\partial {\delta}^2|_{\delta=0}$
(and see Refs.\
\onlinecite{Bishop:2012_honeyJ1-J2,RFB:2013_hcomb_SDVBC,Bishop:2014_honey_XY}
for more details).  The GS phase becomes unstable against
the imposed form of order whenever $1/\chi \rightarrow 0$.  The results are extrapolated to the LSUB$\infty$ limit using the
``leading power-law'' scheme, $\chi^{-1} \rightarrow
x_{0}+x_{1}m^{-\nu}$.  We show LSUB$\infty$ extrapolations based on the
N-II(p) model state in Fig.\ \ref{X_SDVBC}, using this scheme with
LSUB$m$ results $m=\{4,6,8\}$, for various values of $\Delta$.

Figure \ref{X_SDVBC} shows that the locus of the lower critical values
of $\kappa$ at which SDVBC order appears is very insensitive to the
value of $\Delta$ for all $\Delta \gtrsim 0.1$, taking the almost
constant value $\kappa \approx 0.38$.  Very interestingly, the locus
of such SDVBC critical points meets the corresponding locus of
critical values above which N-II(p) order appears (defined as the
corresponding points $\kappa(\Delta)$ at which $M=0$ for the N-II(p)
state, as taken from Fig.\ \ref{M}) at just the value $\Delta \approx
0.1$.  For values $\Delta \lesssim 0.1$, Fig.\ \ref{X_SDVBC} shows
that a region opens up in which both SDVBC and N-II(p) forms of order
seem to coexist over a narrow range of values of $\kappa$, before
N-II(p) order dominates for higher values of $\kappa$.  This ``mixed''
region is denoted as M in the phase diagram shown in Fig.\
\ref{phase}.

We also show in Fig.\ \ref{phase} the regions of stability of the
N($z$) and N(p) phases, taken from Fig.\ \ref{M} as the corresponding
regions in which the respective magnetic order parameters $M$ take
positive values.  

A particularly interesting region in the phase diagram is the
remaining one outside the region of N($z$) stability and
between the two curves $\kappa \approx 0.21$ (below which N(p) order
is stable) and $\kappa \approx 0.38$ (above which SDVBC and/or N-II(p)
order is stable), both curves being almost independent of $\Delta$.

For the $XXX$ model (viz., $\Delta=1$) it remains open as to whether
the GS phase in this region has PVBC order
\cite{Albuquerque:2011_honey,
  Mosadeq:2011_honey,Li:2012_honey_full,Bishop:2012_honeyJ1-J2,RFB:2013_hcomb_SDVBC,Ganesh:2013_honey_J1J2mod-XXX,Zhu:2013_honey_J1J2mod-XXZ}
or is a QSL state
\cite{Clark:2011_honey,Mezzacapo:2012_honey,Yu:2014_honey_J1J2mod-XXZ}.
Of particular interest in this context is a recent DMRG study
\cite{Gong:2013_J1J2mod-XXX} of the $XXX$ model that claimed to find solid
evidence of (weak) PVBC order, in the thermodynamic ($N \rightarrow
\infty$) limit, in the range $0.26 \lesssim \kappa \lesssim 0.35$, but
which excluded (in the same limit) in the range $0.22 \lesssim \kappa
\lesssim 0.26$ immediately above the N\'{e}el-ordered regime both
magnetic (spin) and valence-bond orderings, consistent with a possible
QSL phase.  This result is also in broad agreement with our own
earlier CCM findings,\cite{Bishop:2012_honeyJ1-J2} as we discuss below. 

We have now also tested the stability of the N(p) phase against PVBC
ordering by calculating the corresponding extrapolated inverse
plaquette susceptibility, $\chi^{-1}_{p}$ (see, e.g., Ref.\
\onlinecite{Bishop:2012_honeyJ1-J2}), based on the N(p) model state and
LSUB$m$ results with $m=\{4,6,8\}$.  The corresponding LSUB$\infty$
points at which $\chi^{-1}_{p} \rightarrow 0$ are shown in Fig.\
\ref{phase} by the plus ($+$) symbols.  Based on such results we
tentatively identify the PVBC and QSL regions indicated in Fig.\
\ref{phase}.  The fact that the $+$ symbols for $\Delta \lesssim 0.66$ do
not fall precisely on the lower stability boundary of the N($z$)
regime may be an indication of the error bars associated with the
PVBC boundary points.  These would be reduced by including
higher-order LSUB$m$ results in the extrapolations.  The entire PVBC
and SDVBC regimes would more definitively be confirmed by performing
calculations of $\chi^{-1}_{p}$ and $\chi^{-1}_{d}$ based on
the N-II(p) state, to confirm their respective upper boundaries.  Such
LSUB$\infty$ extrapolations based on the N-II(p) state are technically
more difficult to make, however, and more definitive evidence awaits
higher-order LSUB$m$ calculations.  In their absence we cannot exclude QSL behavior also in the regime between the N($z$)
and SDVBC phases for $\kappa \lesssim 0.66$.

In our earlier CCM analysis of the $XXX$
model\cite{Bishop:2012_honeyJ1-J2} we noted the possibility that the transition from the N(p) phase to the
PVBC phase in that model might be via an intermediate phase, now
identified here as a tentative QSL phase.  Our best estimate then was that such an intermediate phase should be restricted to a
region $\kappa_{c_{1}} < \kappa < \kappa_{c_{1}}'$.  The value of
$\kappa_{c_{1}}$ was very accurately obtained, as
$\kappa_{c_{1}}=0.207(3)$, from the point where N\'{e}el
order vanishes, and is identical to that shown in Fig.\ \ref{phase}
for the boundary between the N(p) and potential QSL phases.  The
accuracy in $\kappa_{c_{1}}$ stems from the shape of the N(p) order
curve shown in Fig.\ \ref{M}, which has an infinite (or very steep)
slope at the point $\kappa =\kappa_{c_{1}}$ where $M \rightarrow 0$.
By contrast $\kappa_{c_{1}}'$ was determined from the
point where $\chi^{-1}_{p} \rightarrow 0$.  Since the slope of the
$\chi^{-1}_{p}(\kappa)$ curve becomes zero (or very
small) at the point where it vanishes, the value of $\kappa_{c_{1}}'$
had a much larger error, and in Ref.\
\onlinecite{Bishop:2012_honeyJ1-J2} we quoted a value $\kappa_{c_{1}}'
\approx 0.24$, with no error estimate.  The present analysis has enabled us to examine the lower phase boundary
of the PVBC phase in more detail, and our best estimate for the $XXX$
model is now $\kappa_{c_{1}}' \approx 0.28(2)$, as indicated in Fig.\
\ref{phase}.  The fact remains that the position of this
phase boundary (i.e., the upper one for the tentative QSL phase) has
the largest uncertainty of all those shown in Fig.\ \ref{phase}, with a similar error along its
length to that quoted above for the case $\Delta=1$.

It is interesting to speculate whether the N($z$)
phase survives to higher values of the spin quantum number $s$ than
the $s=\frac{1}{2}$ case considered here.  In this context we have performed some very preliminary low-order CCM SUB$m$--$m$
calculations with $m \leq 8$ for the spin-1 version of the present
$XX$ model (i.e., with $\Delta=0$).  In the SUB$m$--$m$ scheme one
retains all multispin-flip configurations $\{I\}$ in the expansions of
the CCM correlation operators $S$ and $\tilde{S}$ involving no more
than $m$ single-spin flips spanning a range of $m$ or fewer
contiguous lattice sites.  These
preliminary SUB$m$--$m$ calculations, based on both the N(p) and
N-II(p) model states, exhibit three features: (a) the convergence of both $E/N$ and $M$ with
increasing values of the truncation index $m$ is more rapid than for
the spin-$\frac{1}{2}$ model; (b) the energy curves for the N(p) and
N-II(p) phases meet (or almost meet) near their respective termination points (so far only obtained rather approximately), and
without any perceivable discontinuity in slope, at a value $\kappa
\approx 0.25$; and (c) the corresponding extrapolated order parameters $M$
of both the N(p) and N-II(p) phases appear to go to zero at values of
$\kappa$ very close to the same value $\kappa \approx 0.25$.  These
results alone show rather clearly that if, for the spin-1 model, there
is a phase intermediate between the N(p) and N-II(p) phases, it
can exist in only a very narrow region indeed around $\kappa \approx
0.25$.  Finally, similar calculations based on the
N($z$) state for the spin-1 model show no signs of it providing a stable
GS phase for any values of $\kappa$.  Thus, our preliminary conclusion
is that the stability of the N($z$) phase is restricted to the
spin-$\frac{1}{2}$ system, although more work would be needed to
confirm this.

In conclusion, our CCM analysis gives a coherent picture of the full
$T=0$ GS phase diagram of the model under study.  In particular, we
have identified a candidate QSL regime, in which we have excluded
magnetic or valence-bond forms of order.  It would be of great
interest to use other techniques in order to verify our findings.

We thank the University of Minnesota Supercomputing Institute for the
grant of supercomputing facilities.

\bibliographystyle{apsrev4-1}
\bibliography{bib_general}

\begin{thebibliography}{32}%
\makeatletter
\providecommand \@ifxundefined [1]{%
 \@ifx{#1\undefined}
}%
\providecommand \@ifnum [1]{%
 \ifnum #1\expandafter \@firstoftwo
 \else \expandafter \@secondoftwo
 \fi
}%
\providecommand \@ifx [1]{%
 \ifx #1\expandafter \@firstoftwo
 \else \expandafter \@secondoftwo
 \fi
}%
\providecommand \natexlab [1]{#1}%
\providecommand \enquote  [1]{``#1''}%
\providecommand \bibnamefont  [1]{#1}%
\providecommand \bibfnamefont [1]{#1}%
\providecommand \citenamefont [1]{#1}%
\providecommand \href@noop [0]{\@secondoftwo}%
\providecommand \href [0]{\begingroup \@sanitize@url \@href}%
\providecommand \@href[1]{\@@startlink{#1}\@@href}%
\providecommand \@@href[1]{\endgroup#1\@@endlink}%
\providecommand \@sanitize@url [0]{\catcode `\\12\catcode `\$12\catcode
  `\&12\catcode `\#12\catcode `\^12\catcode `\_12\catcode `\%12\relax}%
\providecommand \@@startlink[1]{}%
\providecommand \@@endlink[0]{}%
\providecommand \url  [0]{\begingroup\@sanitize@url \@url }%
\providecommand \@url [1]{\endgroup\@href {#1}{\urlprefix }}%
\providecommand \urlprefix  [0]{URL }%
\providecommand \Eprint [0]{\href }%
\providecommand \doibase [0]{http://dx.doi.org/}%
\providecommand \selectlanguage [0]{\@gobble}%
\providecommand \bibinfo  [0]{\@secondoftwo}%
\providecommand \bibfield  [0]{\@secondoftwo}%
\providecommand \translation [1]{[#1]}%
\providecommand \BibitemOpen [0]{}%
\providecommand \bibitemStop [0]{}%
\providecommand \bibitemNoStop [0]{.\EOS\space}%
\providecommand \EOS [0]{\spacefactor3000\relax}%
\providecommand \BibitemShut  [1]{\csname bibitem#1\endcsname}%
\let\auto@bib@innerbib\@empty
%</preamble>
\bibitem [{\citenamefont {Rastelli}\ \emph {et~al.}(1979)\citenamefont
  {Rastelli}, \citenamefont {Tassi},\ and\ \citenamefont
  {Reatto}}]{Rastelli:1979_honey}%
  \BibitemOpen
  \bibfield  {author} {\bibinfo {author} {\bibfnamefont {E.}~\bibnamefont
  {Rastelli}}, \bibinfo {author} {\bibfnamefont {A.}~\bibnamefont {Tassi}}, \
  and\ \bibinfo {author} {\bibfnamefont {L.}~\bibnamefont {Reatto}},\
  }\href@noop {} {\bibfield  {journal} {\bibinfo  {journal} {Physica B \& C}\
  }\textbf {\bibinfo {volume} {97}},\ \bibinfo {pages} {1} (\bibinfo {year}
  {1979})}\BibitemShut {NoStop}%
\bibitem [{\citenamefont {Fouet}\ \emph {et~al.}(2001)\citenamefont {Fouet},
  \citenamefont {Sindzingre},\ and\ \citenamefont
  {Lhuillier}}]{Fouet:2001_honey}%
  \BibitemOpen
  \bibfield  {author} {\bibinfo {author} {\bibfnamefont {J.~B.}\ \bibnamefont
  {Fouet}}, \bibinfo {author} {\bibfnamefont {P.}~\bibnamefont {Sindzingre}}, \
  and\ \bibinfo {author} {\bibfnamefont {C.}~\bibnamefont {Lhuillier}},\
  }\href@noop {} {\bibfield  {journal} {\bibinfo  {journal} {Eur. Phys. J. B}\
  }\textbf {\bibinfo {volume} {20}},\ \bibinfo {pages} {241} (\bibinfo {year}
  {2001})}\BibitemShut {NoStop}%
\bibitem [{\citenamefont {Mulder}\ \emph {et~al.}(2010)\citenamefont {Mulder},
  \citenamefont {Ganesh}, \citenamefont {Capriotti},\ and\ \citenamefont
  {Paramekanti}}]{Mulder:2010_honey}%
  \BibitemOpen
  \bibfield  {author} {\bibinfo {author} {\bibfnamefont {A.}~\bibnamefont
  {Mulder}}, \bibinfo {author} {\bibfnamefont {R.}~\bibnamefont {Ganesh}},
  \bibinfo {author} {\bibfnamefont {L.}~\bibnamefont {Capriotti}}, \ and\
  \bibinfo {author} {\bibfnamefont {A.}~\bibnamefont {Paramekanti}},\
  }\href@noop {} {\bibfield  {journal} {\bibinfo  {journal} {Phys. Rev. B}\
  }\textbf {\bibinfo {volume} {81}},\ \bibinfo {pages} {214419} (\bibinfo
  {year} {2010})}\BibitemShut {NoStop}%
\bibitem [{\citenamefont {Ganesh}\ \emph {et~al.}(2011)\citenamefont {Ganesh},
  \citenamefont {Sheng}, \citenamefont {Kim},\ and\ \citenamefont
  {Paramekanti}}]{Ganesh:2011_honey}%
  \BibitemOpen
  \bibfield  {author} {\bibinfo {author} {\bibfnamefont {R.}~\bibnamefont
  {Ganesh}}, \bibinfo {author} {\bibfnamefont {D.~N.}\ \bibnamefont {Sheng}},
  \bibinfo {author} {\bibfnamefont {Y.-J.}\ \bibnamefont {Kim}}, \ and\
  \bibinfo {author} {\bibfnamefont {A.}~\bibnamefont {Paramekanti}},\
  }\href@noop {} {\bibfield  {journal} {\bibinfo  {journal} {Phys. Rev. B}\
  }\textbf {\bibinfo {volume} {83}},\ \bibinfo {pages} {144414} (\bibinfo
  {year} {2011})}\BibitemShut {NoStop}%
\bibitem [{\citenamefont {Clark}\ \emph {et~al.}(2011)\citenamefont {Clark},
  \citenamefont {Abanin},\ and\ \citenamefont {Sondhi}}]{Clark:2011_honey}%
  \BibitemOpen
  \bibfield  {author} {\bibinfo {author} {\bibfnamefont {B.~K.}\ \bibnamefont
  {Clark}}, \bibinfo {author} {\bibfnamefont {D.~A.}\ \bibnamefont {Abanin}}, \
  and\ \bibinfo {author} {\bibfnamefont {S.~L.}\ \bibnamefont {Sondhi}},\
  }\href@noop {} {\bibfield  {journal} {\bibinfo  {journal} {Phys. Rev. Lett.}\
  }\textbf {\bibinfo {volume} {107}},\ \bibinfo {pages} {087204} (\bibinfo
  {year} {2011})}\BibitemShut {NoStop}%
\bibitem [{\citenamefont {Albuquerque}\ \emph {et~al.}(2011)\citenamefont
  {Albuquerque}, \citenamefont {Schwandt}, \citenamefont {Het{\'e}nyi},
  \citenamefont {Capponi}, \citenamefont {Mambrini},\ and\ \citenamefont
  {L{\"{a}}uchli}}]{Albuquerque:2011_honey}%
  \BibitemOpen
  \bibfield  {author} {\bibinfo {author} {\bibfnamefont {A.~F.}\ \bibnamefont
  {Albuquerque}}, \bibinfo {author} {\bibfnamefont {D.}~\bibnamefont
  {Schwandt}}, \bibinfo {author} {\bibfnamefont {B.}~\bibnamefont
  {Het{\'e}nyi}}, \bibinfo {author} {\bibfnamefont {S.}~\bibnamefont
  {Capponi}}, \bibinfo {author} {\bibfnamefont {M.}~\bibnamefont {Mambrini}}, \
  and\ \bibinfo {author} {\bibfnamefont {A.~M.}\ \bibnamefont
  {L{\"{a}}uchli}},\ }\href@noop {} {\bibfield  {journal} {\bibinfo  {journal}
  {Phys. Rev. B}\ }\textbf {\bibinfo {volume} {84}},\ \bibinfo {pages} {024406}
  (\bibinfo {year} {2011})}\BibitemShut {NoStop}%
\bibitem [{\citenamefont {Mosadeq}\ \emph {et~al.}(2011)\citenamefont
  {Mosadeq}, \citenamefont {Shahbazi},\ and\ \citenamefont
  {Jafari}}]{Mosadeq:2011_honey}%
  \BibitemOpen
  \bibfield  {author} {\bibinfo {author} {\bibfnamefont {H.}~\bibnamefont
  {Mosadeq}}, \bibinfo {author} {\bibfnamefont {F.}~\bibnamefont {Shahbazi}}, \
  and\ \bibinfo {author} {\bibfnamefont {S.~A.}\ \bibnamefont {Jafari}},\
  }\href@noop {} {\bibfield  {journal} {\bibinfo  {journal} {J. Phys.: Condens.
  Matter}\ }\textbf {\bibinfo {volume} {23}},\ \bibinfo {pages} {226006}
  (\bibinfo {year} {2011})}\BibitemShut {NoStop}%
\bibitem [{\citenamefont {Oitmaa}\ and\ \citenamefont
  {Singh}(2011)}]{Oitmaa:2011_honey}%
  \BibitemOpen
  \bibfield  {author} {\bibinfo {author} {\bibfnamefont {J.}~\bibnamefont
  {Oitmaa}}\ and\ \bibinfo {author} {\bibfnamefont {R.~R.~P.}\ \bibnamefont
  {Singh}},\ }\href@noop {} {\bibfield  {journal} {\bibinfo  {journal} {Phys.
  Rev. B}\ }\textbf {\bibinfo {volume} {84}},\ \bibinfo {pages} {094424}
  (\bibinfo {year} {2011})}\BibitemShut {NoStop}%
\bibitem [{\citenamefont {Mezzacapo}\ and\ \citenamefont
  {Boninsegni}(2012)}]{Mezzacapo:2012_honey}%
  \BibitemOpen
  \bibfield  {author} {\bibinfo {author} {\bibfnamefont {F.}~\bibnamefont
  {Mezzacapo}}\ and\ \bibinfo {author} {\bibfnamefont {M.}~\bibnamefont
  {Boninsegni}},\ }\href@noop {} {\bibfield  {journal} {\bibinfo  {journal}
  {Phys. Rev. B}\ }\textbf {\bibinfo {volume} {85}},\ \bibinfo {pages}
  {060402(R)} (\bibinfo {year} {2012})}\BibitemShut {NoStop}%
\bibitem [{\citenamefont {Li}\ \emph {et~al.}(2012)\citenamefont {Li},
  \citenamefont {Bishop}, \citenamefont {Farnell},\ and\ \citenamefont
  {Campbell}}]{Li:2012_honey_full}%
  \BibitemOpen
  \bibfield  {author} {\bibinfo {author} {\bibfnamefont {P.~H.~Y.}\
  \bibnamefont {Li}}, \bibinfo {author} {\bibfnamefont {R.~F.}\ \bibnamefont
  {Bishop}}, \bibinfo {author} {\bibfnamefont {D.~J.~J.}\ \bibnamefont
  {Farnell}}, \ and\ \bibinfo {author} {\bibfnamefont {C.~E.}\ \bibnamefont
  {Campbell}},\ }\href@noop {} {\bibfield  {journal} {\bibinfo  {journal}
  {Phys. Rev. B}\ }\textbf {\bibinfo {volume} {86}},\ \bibinfo {pages} {144404}
  (\bibinfo {year} {2012})}\BibitemShut {NoStop}%
\bibitem [{\citenamefont {Bishop}\ \emph {et~al.}(2012)\citenamefont {Bishop},
  \citenamefont {Li}, \citenamefont {Farnell},\ and\ \citenamefont
  {Campbell}}]{Bishop:2012_honeyJ1-J2}%
  \BibitemOpen
  \bibfield  {author} {\bibinfo {author} {\bibfnamefont {R.~F.}\ \bibnamefont
  {Bishop}}, \bibinfo {author} {\bibfnamefont {P.~H.~Y.}\ \bibnamefont {Li}},
  \bibinfo {author} {\bibfnamefont {D.~J.~J.}\ \bibnamefont {Farnell}}, \ and\
  \bibinfo {author} {\bibfnamefont {C.~E.}\ \bibnamefont {Campbell}},\
  }\href@noop {} {\bibfield  {journal} {\bibinfo  {journal} {J. Phys.: Condens.
  Matter}\ }\textbf {\bibinfo {volume} {24}},\ \bibinfo {pages} {236002}
  (\bibinfo {year} {2012})}\BibitemShut {NoStop}%
\bibitem [{\citenamefont {Bishop}\ \emph {et~al.}(2013)\citenamefont {Bishop},
  \citenamefont {Li},\ and\ \citenamefont {Campbell}}]{RFB:2013_hcomb_SDVBC}%
  \BibitemOpen
  \bibfield  {author} {\bibinfo {author} {\bibfnamefont {R.~F.}\ \bibnamefont
  {Bishop}}, \bibinfo {author} {\bibfnamefont {P.~H.~Y.}\ \bibnamefont {Li}}, \
  and\ \bibinfo {author} {\bibfnamefont {C.~E.}\ \bibnamefont {Campbell}},\
  }\href@noop {} {\bibfield  {journal} {\bibinfo  {journal} {J. Phys.: Condens.
  Matter}\ }\textbf {\bibinfo {volume} {25}},\ \bibinfo {pages} {306002}
  (\bibinfo {year} {2013})}\BibitemShut {NoStop}%
\bibitem [{\citenamefont {Ganesh}\ \emph {et~al.}(2013)\citenamefont {Ganesh},
  \citenamefont {van~den Brink},\ and\ \citenamefont
  {Nishimoto}}]{Ganesh:2013_honey_J1J2mod-XXX}%
  \BibitemOpen
  \bibfield  {author} {\bibinfo {author} {\bibfnamefont {R.}~\bibnamefont
  {Ganesh}}, \bibinfo {author} {\bibfnamefont {J.}~\bibnamefont {van~den
  Brink}}, \ and\ \bibinfo {author} {\bibfnamefont {S.}~\bibnamefont
  {Nishimoto}},\ }\href@noop {} {\bibfield  {journal} {\bibinfo  {journal}
  {Phys. Rev. Lett.}\ }\textbf {\bibinfo {volume} {110}},\ \bibinfo {pages}
  {127203} (\bibinfo {year} {2013})}\BibitemShut {NoStop}%
\bibitem [{\citenamefont {Zhu}\ \emph {et~al.}(2013{\natexlab{a}})\citenamefont
  {Zhu}, \citenamefont {Huse},\ and\ \citenamefont
  {White}}]{Zhu:2013_honey_J1J2mod-XXZ}%
  \BibitemOpen
  \bibfield  {author} {\bibinfo {author} {\bibfnamefont {Z.}~\bibnamefont
  {Zhu}}, \bibinfo {author} {\bibfnamefont {D.~A.}\ \bibnamefont {Huse}}, \
  and\ \bibinfo {author} {\bibfnamefont {S.~R.}\ \bibnamefont {White}},\
  }\href@noop {} {\bibfield  {journal} {\bibinfo  {journal} {Phys. Rev. Lett.}\
  }\textbf {\bibinfo {volume} {110}},\ \bibinfo {pages} {127205} (\bibinfo
  {year} {2013}{\natexlab{a}})}\BibitemShut {NoStop}%
\bibitem [{\citenamefont {Gong}\ \emph {et~al.}(2013)\citenamefont {Gong},
  \citenamefont {Sheng}, \citenamefont {Motrunich},\ and\ \citenamefont
  {Fisher}}]{Gong:2013_J1J2mod-XXX}%
  \BibitemOpen
  \bibfield  {author} {\bibinfo {author} {\bibfnamefont {S.-S.}\ \bibnamefont
  {Gong}}, \bibinfo {author} {\bibfnamefont {D.~N.}\ \bibnamefont {Sheng}},
  \bibinfo {author} {\bibfnamefont {O.~I.}\ \bibnamefont {Motrunich}}, \ and\
  \bibinfo {author} {\bibfnamefont {M.~P.~A.}\ \bibnamefont {Fisher}},\
  }\href@noop {} {\bibfield  {journal} {\bibinfo  {journal} {Phys. Rev. B}\
  }\textbf {\bibinfo {volume} {88}},\ \bibinfo {pages} {165138} (\bibinfo
  {year} {2013})}\BibitemShut {NoStop}%
\bibitem [{\citenamefont {Yu}\ \emph {et~al.}(2014)\citenamefont {Yu},
  \citenamefont {Liu}, \citenamefont {Li},\ and\ \citenamefont
  {Zou}}]{Yu:2014_honey_J1J2mod-XXZ}%
  \BibitemOpen
  \bibfield  {author} {\bibinfo {author} {\bibfnamefont {X.-L.}\ \bibnamefont
  {Yu}}, \bibinfo {author} {\bibfnamefont {D.-Y.}\ \bibnamefont {Liu}},
  \bibinfo {author} {\bibfnamefont {P.}~\bibnamefont {Li}}, \ and\ \bibinfo
  {author} {\bibfnamefont {L.-J.}\ \bibnamefont {Zou}},\ }\href@noop {}
  {\bibfield  {journal} {\bibinfo  {journal} {Physica E}\ }\textbf {\bibinfo
  {volume} {59}},\ \bibinfo {pages} {41} (\bibinfo {year} {2014})}\BibitemShut
  {NoStop}%
\bibitem [{\citenamefont {Varney}\ \emph {et~al.}(2011)\citenamefont {Varney},
  \citenamefont {Sun}, \citenamefont {Galitski},\ and\ \citenamefont
  {Rigol}}]{Varney:2011_honey_XY}%
  \BibitemOpen
  \bibfield  {author} {\bibinfo {author} {\bibfnamefont {C.~N.}\ \bibnamefont
  {Varney}}, \bibinfo {author} {\bibfnamefont {K.}~\bibnamefont {Sun}},
  \bibinfo {author} {\bibfnamefont {V.}~\bibnamefont {Galitski}}, \ and\
  \bibinfo {author} {\bibfnamefont {M.}~\bibnamefont {Rigol}},\ }\href@noop {}
  {\bibfield  {journal} {\bibinfo  {journal} {Phys. Rev. Lett.}\ }\textbf
  {\bibinfo {volume} {107}},\ \bibinfo {pages} {077201} (\bibinfo {year}
  {2011})}\BibitemShut {NoStop}%
\bibitem [{\citenamefont {Varney}\ \emph {et~al.}(2012)\citenamefont {Varney},
  \citenamefont {Sun}, \citenamefont {Galitski},\ and\ \citenamefont
  {Rigol}}]{Varney:2012_honey_XY}%
  \BibitemOpen
  \bibfield  {author} {\bibinfo {author} {\bibfnamefont {C.~N.}\ \bibnamefont
  {Varney}}, \bibinfo {author} {\bibfnamefont {K.}~\bibnamefont {Sun}},
  \bibinfo {author} {\bibfnamefont {V.}~\bibnamefont {Galitski}}, \ and\
  \bibinfo {author} {\bibfnamefont {M.}~\bibnamefont {Rigol}},\ }\href@noop {}
  {\bibfield  {journal} {\bibinfo  {journal} {New J.\ Phys.}\ }\textbf
  {\bibinfo {volume} {14}},\ \bibinfo {pages} {115028} (\bibinfo {year}
  {2012})}\BibitemShut {NoStop}%
\bibitem [{\citenamefont {Zhu}\ \emph {et~al.}(2013{\natexlab{b}})\citenamefont
  {Zhu}, \citenamefont {Huse},\ and\ \citenamefont
  {White}}]{Zhu:2013_honey_XY}%
  \BibitemOpen
  \bibfield  {author} {\bibinfo {author} {\bibfnamefont {Z.}~\bibnamefont
  {Zhu}}, \bibinfo {author} {\bibfnamefont {D.~A.}\ \bibnamefont {Huse}}, \
  and\ \bibinfo {author} {\bibfnamefont {S.~R.}\ \bibnamefont {White}},\
  }\href@noop {} {\bibfield  {journal} {\bibinfo  {journal} {Phys. Rev. Lett.}\
  }\textbf {\bibinfo {volume} {111}},\ \bibinfo {pages} {257201} (\bibinfo
  {year} {2013}{\natexlab{b}})}\BibitemShut {NoStop}%
\bibitem [{\citenamefont {Carrasquilla}\ \emph {et~al.}(2013)\citenamefont
  {Carrasquilla}, \citenamefont {Ciolo}, \citenamefont {Becca}, \citenamefont
  {Galitski},\ and\ \citenamefont {Rigol}}]{Carrasquilla:2013_honey_XY}%
  \BibitemOpen
  \bibfield  {author} {\bibinfo {author} {\bibfnamefont {J.}~\bibnamefont
  {Carrasquilla}}, \bibinfo {author} {\bibfnamefont {A.~D.}\ \bibnamefont
  {Ciolo}}, \bibinfo {author} {\bibfnamefont {F.}~\bibnamefont {Becca}},
  \bibinfo {author} {\bibfnamefont {V.}~\bibnamefont {Galitski}}, \ and\
  \bibinfo {author} {\bibfnamefont {M.}~\bibnamefont {Rigol}},\ }\href@noop {}
  {\bibfield  {journal} {\bibinfo  {journal} {Phys. Rev. B}\ }\textbf {\bibinfo
  {volume} {88}},\ \bibinfo {pages} {241109(R)} (\bibinfo {year}
  {2013})}\BibitemShut {NoStop}%
\bibitem [{\citenamefont {Ciolo}\ \emph {et~al.}(2014)\citenamefont {Ciolo},
  \citenamefont {Carrasquilla}, \citenamefont {Becca}, \citenamefont {Rigol},\
  and\ \citenamefont {Galitski}}]{Ciolo:2014_honey_XY}%
  \BibitemOpen
  \bibfield  {author} {\bibinfo {author} {\bibfnamefont {A.~D.}\ \bibnamefont
  {Ciolo}}, \bibinfo {author} {\bibfnamefont {J.}~\bibnamefont {Carrasquilla}},
  \bibinfo {author} {\bibfnamefont {F.}~\bibnamefont {Becca}}, \bibinfo
  {author} {\bibfnamefont {M.}~\bibnamefont {Rigol}}, \ and\ \bibinfo {author}
  {\bibfnamefont {V.}~\bibnamefont {Galitski}},\ }\href@noop {} {\bibfield
  {journal} {\bibinfo  {journal} {Phys. Rev. B}\ }\textbf {\bibinfo {volume}
  {89}},\ \bibinfo {pages} {094413} (\bibinfo {year} {2014})}\BibitemShut
  {NoStop}%
\bibitem [{\citenamefont {Oitmaa}\ and\ \citenamefont
  {Singh}(2014)}]{Oitmaa:2014_honey_XY}%
  \BibitemOpen
  \bibfield  {author} {\bibinfo {author} {\bibfnamefont {J.}~\bibnamefont
  {Oitmaa}}\ and\ \bibinfo {author} {\bibfnamefont {R.~R.}\ \bibnamefont
  {Singh}},\ }\href@noop {} {}\bibinfo {howpublished} {e-print
  arXiv:1403.1905v1 [cond-mat.str-el]} (\bibinfo {year} {2014})\BibitemShut
  {NoStop}%
\bibitem [{\citenamefont {Bishop}\ \emph {et~al.}(2014)\citenamefont {Bishop},
  \citenamefont {Li},\ and\ \citenamefont {Campbell}}]{Bishop:2014_honey_XY}%
  \BibitemOpen
  \bibfield  {author} {\bibinfo {author} {\bibfnamefont {R.~F.}\ \bibnamefont
  {Bishop}}, \bibinfo {author} {\bibfnamefont {P.~H.~Y.}\ \bibnamefont {Li}}, \
  and\ \bibinfo {author} {\bibfnamefont {C.~E.}\ \bibnamefont {Campbell}},\
  }\href@noop {} {}\bibinfo {howpublished} {e-print arXiv:1403.2252v1
  [cond-mat.str-el]} (\bibinfo {year} {2014})\BibitemShut {NoStop}%
\bibitem [{\citenamefont {Bishop}\ and\ \citenamefont
  {K{\"{u}}mmel}(1987)}]{Bishop:1987_ccm}%
  \BibitemOpen
  \bibfield  {author} {\bibinfo {author} {\bibfnamefont {R.~F.}\ \bibnamefont
  {Bishop}}\ and\ \bibinfo {author} {\bibfnamefont {H.~G.}\ \bibnamefont
  {K{\"{u}}mmel}},\ }\href@noop {} {\bibfield  {journal} {\bibinfo  {journal}
  {Phys. Today}\ }\textbf {\bibinfo {volume} {40(3)}},\ \bibinfo {pages} {52}
  (\bibinfo {year} {1987})}\BibitemShut {NoStop}%
\bibitem [{\citenamefont {Arponen}\ and\ \citenamefont
  {Bishop}(1991)}]{Arponen:1991_ccm}%
  \BibitemOpen
  \bibfield  {author} {\bibinfo {author} {\bibfnamefont {J.~S.}\ \bibnamefont
  {Arponen}}\ and\ \bibinfo {author} {\bibfnamefont {R.~F.}\ \bibnamefont
  {Bishop}},\ }\href@noop {} {\bibfield  {journal} {\bibinfo  {journal} {Ann.
  Phys. (N.Y.)}\ }\textbf {\bibinfo {volume} {207}},\ \bibinfo {pages} {171}
  (\bibinfo {year} {1991})}\BibitemShut {NoStop}%
\bibitem [{\citenamefont {Bishop}(1991)}]{Bishop:1991_TheorChimActa_QMBT}%
  \BibitemOpen
  \bibfield  {author} {\bibinfo {author} {\bibfnamefont {R.~F.}\ \bibnamefont
  {Bishop}},\ }\href@noop {} {\bibfield  {journal} {\bibinfo  {journal} {Theor.
  Chim. Acta}\ }\textbf {\bibinfo {volume} {80}},\ \bibinfo {pages} {95}
  (\bibinfo {year} {1991})}\BibitemShut {NoStop}%
\bibitem [{\citenamefont {Bishop}(1998)}]{Bishop:1998_QMBT_coll}%
  \BibitemOpen
  \bibfield  {author} {\bibinfo {author} {\bibfnamefont {R.~F.}\ \bibnamefont
  {Bishop}},\ }in\ \href@noop {} {\emph {\bibinfo {booktitle} {Microscopic
  Quantum Many-Body Theories and Their Applications}}},\ \bibinfo {series and
  number} {Lecture Notes in Physics Vol. 510},\ \bibinfo {editor} {edited by\
  \bibinfo {editor} {\bibfnamefont {J.}~\bibnamefont {Navarro}}\ and\ \bibinfo
  {editor} {\bibfnamefont {A.}~\bibnamefont {Polls}}}\ (\bibinfo  {publisher}
  {Springer-Verlag},\ \bibinfo {address} {Berlin},\ \bibinfo {year} {1998})\
  p.~\bibinfo {pages} {1}\BibitemShut {NoStop}%
\bibitem [{\citenamefont {Bishop}\ \emph {et~al.}(1991)\citenamefont {Bishop},
  \citenamefont {Parkinson},\ and\ \citenamefont
  {Xian}}]{Bishop:1991_XXZ_PRB44}%
  \BibitemOpen
  \bibfield  {author} {\bibinfo {author} {\bibfnamefont {R.~F.}\ \bibnamefont
  {Bishop}}, \bibinfo {author} {\bibfnamefont {J.~B.}\ \bibnamefont
  {Parkinson}}, \ and\ \bibinfo {author} {\bibfnamefont {Y.}~\bibnamefont
  {Xian}},\ }\href@noop {} {\bibfield  {journal} {\bibinfo  {journal} {Phys.
  Rev. B}\ }\textbf {\bibinfo {volume} {44}},\ \bibinfo {pages} {9425}
  (\bibinfo {year} {1991})}\BibitemShut {NoStop}%
\bibitem [{\citenamefont {Zeng}\ \emph {et~al.}(1998)\citenamefont {Zeng},
  \citenamefont {Farnell},\ and\ \citenamefont
  {Bishop}}]{Zeng:1998_SqLatt_TrianLatt}%
  \BibitemOpen
  \bibfield  {author} {\bibinfo {author} {\bibfnamefont {C.}~\bibnamefont
  {Zeng}}, \bibinfo {author} {\bibfnamefont {D.~J.~J.}\ \bibnamefont
  {Farnell}}, \ and\ \bibinfo {author} {\bibfnamefont {R.~F.}\ \bibnamefont
  {Bishop}},\ }\href@noop {} {\bibfield  {journal} {\bibinfo  {journal} {J.
  Stat. Phys.}\ }\textbf {\bibinfo {volume} {90}},\ \bibinfo {pages} {327}
  (\bibinfo {year} {1998})}\BibitemShut {NoStop}%
\bibitem [{\citenamefont {Farnell}\ and\ \citenamefont
  {Bishop}(2004)}]{Fa:2004_QM-coll}%
  \BibitemOpen
  \bibfield  {author} {\bibinfo {author} {\bibfnamefont {D.~J.~J.}\
  \bibnamefont {Farnell}}\ and\ \bibinfo {author} {\bibfnamefont {R.~F.}\
  \bibnamefont {Bishop}},\ }in\ \href@noop {} {\emph {\bibinfo {booktitle}
  {Quantum Magnetism}}},\ \bibinfo {series and number} {Lecture Notes in
  Physics Vol. 645},\ \bibinfo {editor} {edited by\ \bibinfo {editor}
  {\bibfnamefont {U.}~\bibnamefont {Schollw{\"{o}}ck}}, \bibinfo {editor}
  {\bibfnamefont {J.}~\bibnamefont {Richter}}, \bibinfo {editor} {\bibfnamefont
  {D.~J.~J.}\ \bibnamefont {Farnell}}, \ and\ \bibinfo {editor} {\bibfnamefont
  {R.~F.}\ \bibnamefont {Bishop}}}\ (\bibinfo  {publisher} {Springer-Verlag},\
  \bibinfo {address} {Berlin},\ \bibinfo {year} {2004})\ p.\ \bibinfo {pages}
  {307}\BibitemShut {NoStop}%
\bibitem [{\citenamefont {Bishop}\ \emph {et~al.}(2008)\citenamefont {Bishop},
  \citenamefont {Li}, \citenamefont {Darradi}, \citenamefont {Schulenburg},\
  and\ \citenamefont {Richter}}]{Bi:2008_PRB_J1xxzJ2xxz}%
  \BibitemOpen
  \bibfield  {author} {\bibinfo {author} {\bibfnamefont {R.~F.}\ \bibnamefont
  {Bishop}}, \bibinfo {author} {\bibfnamefont {P.~H.~Y.}\ \bibnamefont {Li}},
  \bibinfo {author} {\bibfnamefont {R.}~\bibnamefont {Darradi}}, \bibinfo
  {author} {\bibfnamefont {J.}~\bibnamefont {Schulenburg}}, \ and\ \bibinfo
  {author} {\bibfnamefont {J.}~\bibnamefont {Richter}},\ }\href@noop {}
  {\bibfield  {journal} {\bibinfo  {journal} {Phys. Rev. B}\ }\textbf {\bibinfo
  {volume} {78}},\ \bibinfo {pages} {054412} (\bibinfo {year}
  {2008})}\BibitemShut {NoStop}%
\bibitem [{ccm()}]{ccm}%
  \BibitemOpen
  \href@noop {} {}\bibinfo {note} {We use the program package CCCM of D.~J.~J.
  Farnell and J.~Schulenburg, see
  http://www-e.uni-magdeburg.de/jschulen/ccm/index.html}\BibitemShut {NoStop}%
\end{thebibliography}%

\end{document}